\documentclass[12pt]{article}
\begin{document}

\begin{center}
 {\Large \bf  Separation of variables for a lattice integrable system and the inverse problem.}
\end{center}
\vskip 20pt
\begin{center}
{ Supriya Mukherjee\footnote{E-mail: supriya\_ju@yahoo.com, Present address: Netaji Subhash Engineering College, Dept of Humanities and Science, Kolkata-700 084, India.}, A.Ghose Choudhury$^*$,{A.Roy Chowdhury\footnote{E-mail: arcphy@cal2.vsnl.net.in  } }}\\ 
\end{center}
\begin {center}
High Energy Physics Division, Department of Physics,  Jadavpur University, Calcutta - 700 032, India.\\
$^*$Department of Physics, Surendranath College, 24/2 Mahatma Gandhi Road, Calcutta-700 009, India.  
\end{center}
\vskip 50pt
\begin{center}
{\Large\bf Abstract}\end{center}
 We investigate the relation between the local variables of a discrete integrable lattice system and the corresponding separation variables, derived from the associated spectral curve. In particular, we have shown how the inverse transformation from the separation variables to the discrete lattice variables may be factorised as a sequence of canonical transformations, following the procedure outlined by Kuznetsov. \newpage
\section{ Introduction } Of late there has been a great deal of interest in discrete integrable systems, owing to their numerous interesting properties. One such property which has emerged lately, is the existence of the so called spectral curve associated with an integrable system[1]. The latter in turn is closely connected with the motion of separability. Indeed, it is well known that the technique of separation of variables provides an invaluable tool for the analytic construction of action-angle variables of an integrable system[2]. Furthermore, it may be shown that the locus of points, given by the separation variables, define in fact the spectral curve.
\par On the other hand, the concept of an $r$-matrix algebra marks a major development in theory of nonlinear integrable systems especially in regard to their Hamiltonian nature[3]. It provides an elegant formalism for proving the Poisson involutiveness of the integral of motions for such systems. However, it was only in the eighties that the connection betweenthe $r$-matrix formalism and separation variables of such systems clearly emerged as a result of the efforts of  Komarov [4], Sklyanin [5] and other researchers of the St. Petersburg School. 
\par In this communication, we examine the nature of the relationship between the separation variables of a discrete lattice integrable system and the corresponding local lattice variables. In general,  the nature of the transformation relations between these two sets of variables is rather complicated and an explicit determination of the generating function proves to be rather daunting. The existence of the latter is  essential to gaurantee the canonical nature of the transformation. However, recently Kuznetsov [6] has shown that, not only does the inverse transformation between the separation variables and the local variables exists, but also that these trasformations can even be broken down into a sequence of transformations each with a generating function. 
\par The organisation of the paper is as follows. In section 2, we describe the basic features of the discrete model under consideration. The $r$-matrix algebra of the monodromy matrix is explicitly solved, as also the leading order structures of its elements. In section 3, we introduce the spectral curve and separation representation for the elements of the monodromy matrix. In section 4, which constitutes the main body of the paper, we derive explicitly the inverse separating map between the separation variables and the local lattice variables defining our system. The canonical nature of this mapping is displayed by working out the corresponding generating function. 
\section{ Formulation } We begin by considering a discrete integrable system which is described by a local Lax operator $ l_n ( u ) $ of the form 
$$  l_n ( u ) = \left (\begin{array}{cc} u + q_n p_n  & -p_n \\ q_{n} ^2 p_n & u-q_np_n  \end {array}\right) \eqno( 2.1) $$
Here u is the complex spectral parameter and $ q_n $ and $ p_n $are the nonlinear variables defined at the nth lattice points. They satisfy the canonical Poisson brackets  $$ \{ q_k , q_j \} = \{ p_k , p_j \} = 0   \;\;for\;\; all \;\;  k, j = 1,2,........, n\;\; and  \;\;  \{ q_k , p_ j \} = \delta _{ kj } \eqno( 2.2 ) $$ 
 It is a matter of simple computation to show that 
$$ \{ l_n (u )  \stackrel{\otimes}{,} l_m ( v ) \}= [ r ( u-v ), l_n( u ) \otimes l_m( v ) ] \delta _{ mn } \eqno     (2.3)  $$ where $ r( u-v) $ represents the classical r-matrix given by 
$$ r( u ) = \frac { P }{ u } = \frac { 1 }{ u  } \left (\begin{array}{cccc} 1 & 0&0&0 \\ 0&0&1&0 \\0&1&0&0 \\ 0&0&0&1  \end {array}\right),$$ with $P$ being the permutation matrix satisfying $ P^2 =I.$ 
We define the monodromy matrix $T_N ( u ) $ as the ordered product of $ l_n ( u ) 's$ over the entire lattice. That is,
 $$ T_N ( u ) = \prod_{ n = N  }^1 l_n (  u )  = l_N ( u ) l_ {N-1} ( u ) ...............l_2 ( u ) l_1( u ) \eqno (2.4) $$
Upon using ( 2.3 ) it is easy to deduce that 
$$ \{ T_N ( u )  \stackrel{\otimes}{,}  T_N ( v ) \} = [ r ( u-v ), T_N( u ) \otimes T_N( v ) ]  \eqno (2.5)$$
Consequently we infer that structurally the Poisson algebra of  $l_n ( u ) $ and $ T_N ( u ) $ are identical. In the literature on nonlinear integrable systems equation (2.5) is usually referred to as the quadratic $r $- matrix algebra. In case of $2\times 2$ local Lax operators the monodromy matrix is itself a $2\times2$ matrix which may formally be written as,
$$ T_N ( u ) = \prod_{ n = N  }^1 l_n (  u )  \equiv \left (\begin{array}{cc} A_N (u) & B_N ( u ) \\  C_N (u) & D_N (u) \end {array}\right) \eqno(2.6 ).$$ 
The Sklyanin algebra ( 2.5) then takes the following explicit form, namely 
$$ \{ A_N(u),A_N(v)\} = \{ B_N(u), B_N(v)\} = \{ C_N(u), C_N(v)\}=\{ D_N(u), D_N(v)\}=0, $$
  $$  \{C_N(u), A_N(v)\}= \frac{1}{ u-v } \{A_N(u)C_N(v)- A_N(v)C_N(u)\},$$
  $$ \{A_N(u), D_N(v) \}=\frac{1}{  u-v } \{  C_N ( u) B_N(v)-  C_N ( v) B_N(u)\},$$
  $$ \{B_N(u), A_N(v)\}=\frac{ 1 }{ u-v }  \{B_N(u)A_N(v)- B_N(v)A_N(u)\},$$
$$ \{ C_N (u) , D_N(v) \} = \frac{1 }{ u-v } \{ C_N(u) D_N(v) - C_N(v) D_N(u)\},  $$
$$ \{B_N(u), D_N(v)\} = \frac{1}{ u-v } \{D_N(u) B_N(v)-D_N(v)B_N(u)\}.\eqno( 2.7 )$$
From the definition of monodromy operator in (2.6) and the form of the local Lax operator as given by (2.1) by explicit calculation obtain the leading order terms of the matrix elements $A_N ( u) , B_N(u) $ etc . These are as follows 
$$ A_N(u) = u^N + (  \Sigma _ { i=1 } ^ N q_i p_i ) u^{ N-1} +........................\hskip 175pt \eqno (2.8a) $$
$$ B_N( u ) = ( - \Sigma _ { i=1 } ^ N p_i ) u^ { N-1}+.................................. \hskip175pt \eqno( 2.8b) $$ 
$$ C_N( u ) = (\Sigma _ { i=1 } ^ N q_i ^2 p_i ) u^ { N-1}+.................................. \hskip175pt \eqno( 2.8c) $$
$$ D_N( u ) = u_N  - ( \Sigma _ { i=1 } ^ N q_i p_i ) u^{N -1}  +.................................. \hskip175pt \eqno( 2.8d) $$ 
Note that $ det $$l_n (u) = u^2 $ so that $ det T_N (u) = u^{2N} $. Now if at each lattice site an inhomogenity parameter $ c_n $ is introduced, then the local Lax operator assumes the following form: 
$$ l_n ( u-c_n  ) = \left (\begin{array}{cc} u -  c_n + q_ n p_n  & -p_n\\ q_{n} ^2 p_ n & u - c_n - q_n p_n \end {array}\right), \eqno( 2.9) $$ 
and consequently the monodromy matrix 
$$T_N ( u, \{c_n \} ) = \left (\begin{array}{cc} A_N (u) & B_N ( u ) \\  C_N (u) & D_N (u) \end {array}\right) $$
now involves the inhomogenity parameters $ \{ c_n \}_{n=1} ^N $ . Its elements then are as given below 
$$ A_N(u ,\{c_n \}  ) = u^N + (  \Sigma _ { i=1 } ^ N q_i p_i - \Sigma_{ i=1 }^N c_i ) u^{ N-1} +........................\hskip 175pt \eqno (2.10a) $$
$$ B_N( u,\{c_n \}  ) = ( - \Sigma _ { i=1 } ^ N p_i ) u^ { N-1}+.................................. \hskip175pt \eqno( 2.10b) $$ 
$$ C_N( u,\{c_n \}  ) = (\Sigma _ { i=1 } ^ N q_i ^2 p_i ) u^ { N-1}+.................................. \hskip175pt \eqno( 2.108c) $$
$$ D_N( u,\{c_n \}  ) = u_N  - ( \Sigma _ { i=1 } ^ N q_i p_i +\Sigma_{ i=1 }^N c_i  ) u^{N -1}  +.................................. \hskip175pt \eqno( 2.10d). $$
Thus in general $ A_N , D_N $  are polynomials of degree N, while $B_N $and $C_N $ are polynomials of degree N-1 in the spectral parameter. Furthermore, in this case 
 $$det\;\; l_n ( u-c_n ) = (u - c_n ) ^2 $$so that $$det\;\;  T_N (u,\{ c_n\} ) =  \prod_{ n = 1  }^N ( u - c_n )^2    \eqno (2.11) $$
 The involutive, independent integrals of the system are now obtained as the coefficients of trace of the monodromy matrix. That is,  
$$ tr T_N(u,\{ c_n\}) = A_N( u,\{ c_n\}) + D_N ( u,\{ c_n\} )  \hskip200pt$$$$=2 u^N  +H_1 u^{N-1}  + H_2 u^{N-2} + ......... + H_{N-1} u+ H_N\eqno( 2.12 ) $$
For a three particle system i.e N=3 
$$  H_1 = -2( c_1 +c_2+c_3)  \hskip350pt $$$$ H_2=  4( c_3c_1 + _1c_2 + c_2c_3 ) + 2 ( k_3k_2+k_2k_1+k_1k_3 )  - k_3\rho _3p_2-k_3\rho _3p_1-k_2\rho _2p_3-k_2\rho _2p_1-k_1\rho _1p_3 -k_1\rho _1p_2 $$
 Note that throughout this paper we shall assume periodic boundary conditions i.e. $ q_{ n+N } = q_n \;\;and \;\; p_{n+N } = p_n $
\section { Spectral curve and Separation representation } In case of Liouville integrable systems, the technique of separation of variables play a vital role in construction of analytic expressions for the action angle variables. However it was only in the early eighties that the connection between separation variables and the r-matrix formalism of the quantum inverse scattering method became evident. The fact that for $2\times 2$ Lax matrices, the separation variables are given by the zeros of the off diagonal of $ T_N ( u) $ was first observed by Komarov [4]. The technique was investigated in great detail by Sklyanin [5] and was applied to several models, such as the Goryachev-Chaplygin top, the Neuman system [8] etc.  Indeed, its subsequent application to the O(4) Kowalevski top, by Kuznetsov, represents the culmination in a sense, of this remarkable technique[9] . 
\par We shall briefly outline the connection between separation variables and the elements of the so called spectral curve, which is explained below. It is well known that, an integrable Hamiltonian system possessing  N independent integrals of motion in involution, is separable, if there exists a set of Darboux co-ordinates say  $( u_i, v_i )_{ i=1}^N $ which satisfy a set of equations of the form 
$$ f ( u_i, v_i, I_1, I_2, ................... , I_N ) = 0 \;\;\; ( i=1,2,........, N ) \eqno( 3.1 ).$$ These equations arise from the equation of an associated spectral curve $ \Gamma $ defined by the following equation  
$$ det ( zI - T_N (u) ) = z^2 - P(u) z + Q(u) =0, \eqno (3.2)  $$
 on the ( u, z ) plane. Here $P(u)$ and $Q (u)$ are given by $ tr ( T_N(u) ) $ and $ det (T_N(u) ) $ respectively. 
In  our case we note that $ B_N ( u ) $ i.e the upper off diagonal element of $ T_N(u) $ is a polynomial of degree (N-1), so that it can have at most ( N-1) distinct zeros, which we shall denote by $ u_i\;\;( i=1,2,......., N-1) $. One may show that there exists a system of differential equations for these zeros. It is easy to see that at the zeros of $ B_N (u) $, viz. $ u_i \; ( i= 1,2,...,N-1 ) $ the monodromy matrix becomes triangular, i.e, 
$$ T_N ( u_i ) = \left (\begin{array}{cc} A_N (u_i) & 0 \\  C_N (u_i) & D_N (u_i) \end {array}\right) \eqno( 3.3 )$$  
and therefore its eigen-values are the diagonal elements $ A_N ( u_i ) $ and  $ D_N(u_i) $ repectively. If $ z_i \equiv  A_N(u_i) , \;\; i=1,2,........,N-1 $, then it is obvious that the set of points $(u_i,z_i )_{ i=1 }^N  $ of the (u-z) plane must lie on the spectral curve $ \Gamma _i $
$$  z_i^2 - P( u_i ) z_i  + Q( u_i  ) =0  \;\; i=1,2,........,N-1. \eqno( 3.4 )$$ 
For the remaining part of this communication we shall set  $ z_i = e^{ -v_i } $ and shall  show that the pair  $ ( u_i , v_i ) $ constitute a set of Darboux co-ordinates satisfying the canonical brackets 
$$ \{ u_i , u_j \} = \{ v_i , v_j \} = 0  ,\;\;\;  \{ u_i , v_ j \} = \delta _{ ij } \;\; i,j = 1,2,......., N-1.\eqno( 3.5 ) $$
Let us further define two quantities as follows:
$$  u_N \equiv \frac{ \Sigma_{ i=1 }^N q_ip_i - c_i  }{ - \Sigma _{ i=1 }^N p_i } ,\;\;\; v_N \equiv - \Sigma_{ i = 1 }^Np_i ,   $$ satisfying $ \{ u_N, v_N \} =1 $ 
Thus altogether  we have a total of N pairs of Darboux co-ordinates $ (u_i, v_i )_{ i=1 }^N $. 
Having defined the above separation variables we next proceed to the construction of a representation of the quadratic $r$-matrix algebra (2.5) in terms of these separation variables. This is easily accomplished by using Lagrange interpolation. For instance, we may express  $ B_N(u) $ in terms of its zeros $ u_1, \;u_2,\; ..........u_{ N-1} $ as 
$$ B_N(u) = v_N \prod_{ n =  1 }^{N-1} ( u-u_i ). \eqno (3.7) $$
Furthermore, assuming these zeros to be mutually distinct we may express any polynomial F(u) of degree less than (N-1) in the following mannner 
$$  \frac { F(u) }{ B_N(u)} =  \Sigma_{ i=1 }^{(N-1)} \frac { F( u_i )}{ B'_N (u_i) (u-u_i )},\eqno(3.8) $$
while if F(u) has degree $\geq $ (N-1), then 
$$  \frac { F(u) }{ B_N(u)} = \Sigma_{ i=1 }^{(N-1)} \frac { F( u_i )}{ B'_N  (u_i) (u-u_i )} + G(u) \eqno(3.9) $$ 
where G(u) is a polynomial of appropriate degree. In the present case since $ A_N (u)$ and $D_N (u) $ are polynomials of degree N, we have the following representations for them,
$$ A_N(u) = B_N(u)  [ \frac { u+ ( u_Nv_N + \Sigma _{ i=1 }^{(N-1)} u_i )}{ v_N } +  \Sigma_{ i=1}^{ N-1} \frac{ e^{ -v_i}}{B'_N  (u_i) ( u-u_i)} ]  \eqno (3.10) $$
$$ D_N(u) = B_N(u)  [ \frac { u- ( u_Nv_N - \Sigma _{ i=1 }^{(N-1)} u_i  + 2 \Sigma_{ i=1}^N c_i )}{ v_N } +  \Sigma_{ i=1}^{ N-1} \frac{ det T_N (u_i) e^{ v_i }}{B'_N  (u_i) ( u-u_i)} ]  \eqno (3.11) $$
since $$det T_N(u_i) = A_N(u_i) D_N(u_i )= e^{ -v_i } D_N(u_i) \eqno(3.12)$$
The representation for $ C_N (u) $ follows from the relation 
$$ det T_N(u)  = A_N(u) D_N(u) - B_N(u) C_N(u) $$ so that 
$$ C_N(u) = \frac {A_N(u) D_N(u) -det T_N(u)}{ B_N(u)} \eqno(3.13) $$
with $ A_N(u) , D_N(u),\;\; and \;\; B_N(u) $ as given by the relations (3.10) , ( 3.11), and (3.7) while 
 $$ detT_N= \prod_{ n = N  }^1 det l_n ( u-c_n ) =   \prod_{ n = 1  }^N ( u - c_n )^2.$$
We shall now proceed to show that the above representations for the elements of $ T_N(u) $ does indeed reproduce the quadratic r-matrix algebra. For example, 
$$ \{ B_N(u) , A_N(v) \} = \{ B_N(u) , B_N(v)  [ \frac { v+ ( u_Nv_N + \Sigma _{ i=1 }^{(N-1)} u_i )}{ v_N} +  \Sigma_{ i=1}^{ N-1} \frac{ e^{ -v_i}}{B'_N  (u_i) ( v-u_i)} ]  \}$$
$$\;\;\;\; \;\;\;\; \;\;\;\;\;\;\;\;\;\;\;\;\:\;\;=  B_N(v) \{ B_N(u) ,       \frac { v+ ( u_Nv_N + \Sigma _{ i=1 }^{(N-1)} u_i )}{ v_N } +  \Sigma_{ i=1}^{ N-1} \frac{ e^{ -v_i}}{B'_N  (u_i) ( v-u_i)}   \} $$
since $ \{ B_N(u) , B_N(v) \} = 0 $.
$$ = B_N(v) \{ v_N, u_N \} \prod_{ n = 1  }^{N-1} (u-u_i) + B_N(v) \Sigma_{ i=1}^{ N}-1v_N \prod_{k=1}^{i=1}  ( u-u_k) \{ -u_i, e^{ v_i} \} \frac{1}{ B'_N  (u_i) ( v-u_i)}\prod_{k=i+1}^{N-1}  ( u-u_k) $$
$$ = - \frac {B_N(v)B_N(u)}{ v_N} + \frac { B_N(v)B_N(u) }{ u-v } \Sigma _{i=1}^{ N-1} \frac { e^{-v_i}}{ B'_N(u_i)} ( \frac {1}{v-u_i}- \frac{ 1}{u-u_i} ) \hskip150pt $$
$$ = - \frac {B_N(u)B_N(v)}{ v_N} + \frac { B_N(u)B_N(v) }{ u-v }\large( \frac { A_N(v) }{ B_N(v)} -\frac{  v+  u_Nv_N+ \Sigma _{ i=1 }^{(N-1)} u_i }{ v_N}  -   \frac { A_N(u) }{ B_N(u)} +\frac {u+  u_Nv_N + \Sigma _{ i=1 }^{(N-1)} u_i ) }{ v_N }\large)$$
$$ \{ B_N(u), A_N(v) \} = \frac { 1}{ u-v } ( B_N(u) A_N(v) - B_N(v) A_N(u) ).\hskip150pt $$
 It is also straight forward to show that $det T_N (u)$ is a Casimir, i.e, its Poisson brackets with the elements of  $T_N ( u )$ vanish.
 \section { Factorized separation } From the above discussion it is obvious that we have at our disposal two sets of local variables-- the first set consisting of  $( q_i, p_i )_{i=1} ^N $, in terms of which the local Lax matrices are represented  and a second  set consisting of the separation variables $ ( u_i, v_i )_{ i=1} ^N $. The two sets of variables being connected by the separation map $ S_i $
given by 
$$ B_N (u_i) = 0 , \;\; e^{ v_i }= A_N (u_i) \;\; i=1,2 ...........,(N-1) \eqno ( 4.1a) $$
while $$ u_N = \frac{ \Sigma _{ i=1}^N ( q_i p_i - c_i ) }{ -\Sigma _{ i=1}^Np_i  } , \;\; 
v_N = -\Sigma _{ i=1}^Np_i .\eqno(4.1b)  $$
The concept of factorized separation has recently been introduced by Kuznetsov in [6] to obtain an inverse of the separation map. Indeed, the mapping  $ S_N $ between the original local variables $ (p_i , q_i ) $ and the separation variables is in effect a rather complicated canonical tranformation and to obtain its generating function is a tedious task. The same also holds for the inverse mapping which relates the separation variables to the local variables appearing in the Lax operators. In [6] Kuznetsov made the remarkable observation that not only does the inverse mapping $ S_N^{ -1 }$ exist but that it could also be factorized as a chain of more simpler, at least in principle, canonical transforms, thus leading to the solution of the inverse problem.
\par We note that the N-site monodromy matrix may also be expressed as 
$$  T_N (u) = l_N(u-c_N) T_{ N-1} (u) =  \left (\begin{array}{cc} u -  c_N + q_N p_N  & -p_N \\ q_{N} ^2 p_N & u - c_N - q_Np_N  \end {array}\right) T_{ N-1}(u), \eqno(4.2) $$
while $$ T_{N-1} (u) = l_N^{-1}(u-c_N) T_  N  (u) =   \frac{ Adj l_N (u-_n) }{ det l_N (u-c_N)}  T_N(u). \eqno(4.3) $$ 
If we formally denote $$ T_ { N-1 } (u) =  \left (\begin{array}{cc} A_{N-1} (u) & B_{N-1}(u)  \\  C_{ N-1}  (u) & D_{N-1} (u  ) \end {array}\right), \eqno(4.4) $$  
then it is obvious that $ B_{N-1}(u) $ is a plynomial of degree (N-2) while $A_{N-1} (u) $ and $ D_{N-1} (u) $ are polynomials of degree (N-1) in $u$. Moreover, by analogy with their earlier counterparts, let  $\tilde u_j \;\;(j=1,2,.....,(N-2)) $ denote the zeros of $B_{N-1} $, so that 
$$ B_{N-1} (u)(\tilde u_j )=0  ,\;\;\  j=1,2,......,(N-2) \eqno (4.5) $$ and let  $$ e^{ - v_i } = A_{N-1}(\tilde u_j) ;\;\;\;\ j=1,2,......,(N-2) \eqno (4.6) $$ 
while $$\tilde u_{N-1} = \frac{ \Sigma _{ i=1}^{N-1} ( q_i p_i - c_i ) }{ -\Sigma _{ i=1}^{N-1}p_i  } , \;\; 
\tilde v_{N-1} = -\Sigma _{ i=1}^{  N- 1}p_i \eqno(4.7) $$ 
As before  $ A_{N-1} (u) $ and $ D_{N-1}(u) $ may in view of (4.5) and (4.6) be expressed by Lagrange interpolation formulae in the undermentioned  forms: 
$$ A_{N-1}(u) = B_{N-1}(u)  [ \frac { u+ ( \tilde u_N\tilde v_N + \Sigma _{ j=1 }^{(N-2)} \tilde u_j )}{ \tilde v_N } +  \Sigma_{ j=1}^{ N-2} \frac{ e^{ -\tilde v_j}}{B'_N  (\tilde u_j) ( u-\tilde u_j)} ]  \eqno (4.8a)  $$
$$ D_{N-1}(u) = B_{N-1}(u)  [ \frac { u- ( \tilde u_N\tilde v_N - \Sigma _{ j=1 }^{(N-2)} \tilde u_j + 2 \Sigma_{ i=1}^N c_i )}{ \tilde v_N } +  \Sigma_{ j=1}^{ N-2} \frac{ det T_{N-1} (\tilde u_j) e^{ \tilde v_j }}{B'_{N-1}  (\tilde u_j) ( u-\tilde u_j)} ]  \eqno (4.8b) $$
while  $$ B_{N-1} (u) = \tilde v_{N-1}   \prod_{j=  1}^{N-1}( u-\tilde  u_j). \eqno (4.9) $$
Let $ X_N $ denote the mapping from the separation variables $ u_i, v_i ( i=1,2,.....,(N-1)) $ to the new separation variables $\tilde u_j , \tilde v_j \;\;( j=1,2,.....,(N-2)  $ which parameterize $T_{N-1}(u) $ and into the pair of local variables  $ q_N $ and $ p_N $. From (4.3), since $ \det\; l_N(u-c_N) = (u-c_N)^2 $, we see that  $T_{N-1} (u) =   \frac{ Adj\; l_N (u-c_N) }{   (u-c_N)^2 }  T_N(u) $ has a second order pole at $u=c_N$. As $ T_{N-1 }(u) $ does not have any singularities, consequently upon equating the residue at $ u=c_N $ to zero, we have 
$$ \lim_ { u\longrightarrow c_N } \frac{d}{du}[ Adj \;l_N (u-c_N) T_N(u) ] =0, \eqno (4.10) $$
which leads to the following relations: 
$$\left (\begin{array}{cc} - q_N p_N  &  p_N \\ -q_{N} ^2 p_N & q_Np_N  \end {array}\right)=\frac{1}{\bigtriangleup } \left (\begin{array}{cc} A_N (c_N)  & B_N (c_N)  \\ C_{N}(c_N)   & D_N(c_N)  \end {array}\right)  \left (\begin{array}{cc} \frac{d}{du}D_N(u) & -\frac{d}{du}B_N(u) \\-\frac{d}{du}C_N(u) &\frac{d}{du}A_N(u) \end {array}\right) _{u=c_N }\eqno(4.11) $$
where $$\bigtriangleup = [\frac{d}{du}A_N(u)  \frac{d}{du}D_N(u) -\frac{d}{du}B_N (u) \frac{d}{du}C_N(u) ] _{u=c_N}.  \eqno (4.12) $$
Hence $$ p_N = \frac{1} { \bigtriangleup } [ -A_N(c_N) \frac{d}{du}B_N(u)+ B_N(c_N)\frac{d}{du} A_N(u)]_{u=c_N} \eqno ( 4.13 )$$and  $$ q_N=[ \frac{-C_N(u) \frac{d}{du}B_N(u)+ D_N(u)\frac{d}{du} A_N(u)}{-A_N(u) \frac{d}{du}B_N(u)+ B_N(u)\frac{d}{du} A_N(u)}] _{u=c_N}\eqno(4.14) $$
Equations (4.13 - 4.14) determine the Nth local variables in terms of the separation variables $ (u_i, v_i )|_{i=1} ^N .$ To determine the remaining local variables we adopt a recursive procedure. To this end we shall first consider the mapping $$ X_N : ( \vec u, \vec v )\longmapsto (\vec{\tilde    u},  \vec {\tilde   v }|q_N,p_N ), \eqno( 4.15) $$where  $\vec u= ( u_1,u_2, ...........u_N ) etc. $ 
Next if we consider the mapping $ X_{N-1} $ given by, 
$$ X_{N-1} : (  \vec{\tilde u}, \vec {\tilde  v}| q_N,p_N )\longmapsto (\vec {\tilde {\tilde   u}}, \vec {\tilde  {\tilde   v }}|q_N,p_N ; q_{N-1},p_{N-1}  ) \eqno( 4.16) $$
then it is obvious that the composition of these two mappings, $$ X_{N-1} \circ X_N ) : ( \vec u, \vec v ) \longmapsto (\vec {\tilde {\tilde   u}}, \vec {  \tilde  {\tilde   v }}|q_N,p_N ; q_{N-1},p_{N-1}  ) \eqno( 4.17) $$
Therefore the composition of N such mappings viz  $$ X_ 1   \circ X_2\circ ............\circ  X_{N-1} \circ X_N    : ( \vec u, \vec v ) \longmapsto (\vec  q, \vec p ) \eqno ( 4.18) $$ leads to a transformation from the separation variables $ (u_i, v_i )|_{i=1} ^N $ to the local variables $ ( q_i, p_i )_{i=1} ^N  $ , which is precisely the definition of $ S_N^{-1} $i.e,  $$ S_N^{-1}=  X_ 1   \circ X_2\circ ............\circ  X_{N-1} \circ X_N  \eqno(4.19) $$
Equation ( 4.19) represents the separation factorization chain of the inverse mapping from the separation variables to the local variables occuring in the Lax operators. 
\par In the present case the mapping $ X_N $ will be explicitly derived in the following manner[6].
From equation (4.2) we obtain 
$$ B_N(u) = ( u-c_N + q_N p_N ) B_{N-1} (u) - p_N D_{N-1} (u) \eqno(4.20) $$ 
If this equation is evaluated at  $ u=\tilde u_j   ( j=1,2,....., N-2) $ i.e, at the zeros of $ B_{N-1} ( u) $ then one obtains the following relation 
$$ B_N(\tilde u_j ) = - p_N D_{N-1} ( \tilde u_j ) \;\;\; j= 1,2 ,........ , N-2 \hskip250pt \eqno ( 4.21)   $$
$$ B_N(\tilde u_j ) = - p_N \{det T_{N-1} ( \tilde u_j ) e^{\tilde v_j} \}\;\; j= 1,2 ,........ , N-2 \hskip250pt  \eqno ( 4.22) $$ $$ = -p_N \prod_{ k = 1  }^{N-1} ( \tilde u_j - c_k )^2 e^{\tilde v_j } \hskip250pt$$
Again from (4.3) we obtain 
$$ B_{N-1} ( u) = \frac{1} { det\; l_N ( u-c_N )} [ ( u-c_N - q_Np_N ) B_N(u) + p_N D_N(u) ] \eqno(4.23) $$
Its evaluation at $ u=u_i $ gives 
$$ ( u_i - c_N )^2 B_{N-1} (u_i) = p_N D_N(u_i) \;\;\; i=1,2,.....,N-1 \eqno( 4.24) $$ 
Substituting expressions for $ B_N (u) $ and $ B_{N-1} (u) $ from (3.7), and (4.9) respetively, and $ D_{N-1} (u) $ from (4.8b) into (4.20) and equating the leading order terms, we obtain from the coefficient of $ u^{N-1}, $ 
$$ v_N = \tilde v_{N-1} - p_N .\eqno(4.25)$$ 
However,  it is also follows from the definitions of $ ( u_N, v_N)$ and $( \tilde u_{N-1} , \tilde v_{N-1}) $ as given in (4.1b) and (4.7) repectively that $$ u_Nv_N = \Sigma_{i=1} ^N (q_ip_i - c_i ) \;\; and \;\; \tilde u_{N-1}   \tilde v_{N-1} = \Sigma_{i=1} ^{N-1} (q_ip_i - c_i ), \eqno(4.26)  $$
so that their difference i.e, 
$$ u_N v_N -  \tilde u_{N-1}   \tilde v_{n-1} = q_N p_N -c_N .\eqno( 4.27 ) $$
 Now from (4.25) and (4.27) we may solve for $ v_N $ and $\tilde v_N $ to obtain, 
$$  v_N = \frac{ p_N \tilde u_{N-1} + q_N p_N -c_N }{ u_N  - \tilde u_{N-1}} \eqno( 4.28a )$$
$$  \tilde v_{N-1} = \frac{ p_N u_N + q_N p_N -c_N  }{u_N  - \tilde u_{N-1} } \eqno ( 4.28b)$$
Equating the coefficients of $ u^{N-2} $ in (4.20) we obtain 
$$ -v_N \Sigma _{ i=1}^{N-1} = \tilde v_{N-1} \{ -\Sigma_{j=1} ^{N-2} \tilde u_j + (q_N p_N -c_N ) \} +p_N ( \tilde u_N \tilde v_N + 2 \Sigma _{i=1 } ^{N-1} c_i ) $$
 leading to 
$$ p_N = \frac {- v_N \Sigma _{i=1} ^{N-1} u_i+ \tilde v_{N-1}  (\Sigma_{j=1}^{N-2} \tilde u_j +   c_N )}{ ( \tilde u_N \tilde v_N + 2\Sigma _{i=1}^{N-1} c_i  +q_N \tilde v_{N-1} )}. \eqno ( 4.29) $$
Substituting the expressions for $ v_N $ and $ \tilde v_{N-1} $ from (4.28a,b) into (4.29) yeilds the following quadratic expressions for  $p_N  ( q_N , \vec u,\tilde  {\vec u }, \{ c_i\} ) $ namely, 
$$ p_N ^ 2 (q_N + u_N ) ( q_N + \tilde u_{N-1} ) + p_N [ (q_N + \tilde u_{N-1}  )(\Sigma_{ i=1}^{N-1}  u_i -   c_N  ) - (q_N + u_N ) ( \Sigma_{j=1}^{N-2} \tilde u_j +   c_N )  $$ $$+ 2\Sigma _{i=1}^{N-1} c_i    ( u_N - \tilde u_{N-1} )] + c_N [ c_N +\Sigma_{j=1}^{N-2} \tilde u_j -\Sigma_{i=1}^{N-1}   u_i ] =0. \eqno ( 4.30) $$ 
Incidentally,  if we set  $c_i =0 $ for all $ i= 1,2,....,N ,$ then (4.30) leads to the following linear expressions for $ p_N $ 
$$ p_N = [ \frac{ \Sigma_{j=1}^{N-2} \tilde u_j}{ q_N +\tilde u_{N-1} } - \frac { \Sigma_{i=1}^{N-1}   u_i }{ q_N+u_N } ] \eqno(4.31) $$
On the other hand from (4.22) we get 
$$ e ^{ -\tilde v_j} = -\frac {p_N  \prod_{ k = 1  }^{N-1} ( \tilde u_j - c_k )^2 }{ B_N ( \tilde u_j ) } =-\frac{ p_N \prod_{ k = 1  }^{N-1} ( \tilde u_j - c_k )^2}{ v_N \prod_{ k = 1  }^{N-1} ( \tilde u_j - u_k )   }, \;\;\;\;\; j=1,2,....,N-2 \eqno (4.32) $$ 
Upon substituting $ v_N $ from (4.28a ) we get 
$$ e^{-\tilde v_j } = - \frac{ p_N (u_N - \tilde u_{N-1}) }{p_N ( q_N+\tilde u_{N-1} ) - c_N    } \prod_{ k = 1  }^{N-1} \frac{ ( \tilde u_j - c_k )^2}{ (\tilde u_j -u_k) } \;\;\;\; j=1,2,............., N-2 \eqno (4.33) $$
Finally from (4.24) we have upon using the expressions for $ B_{N-1} (u_i) $ and $ D_N(u_i)  $ obtained from (4.9) and (3.11) repectively, the following relation, 
$$ e ^{ -v_i} =  \frac{ p_N }{\tilde v_{N-1}( u_i -c_N )^2  }\frac{\prod_{ m = 1  }^{N-1} ( \tilde u_i - c_m )^2}{  \prod_{ j = 1  }^{N-2} ( u_i - \tilde u_j )   }, \;\;\;\;\; i=1,2,....,N-1 \eqno (4.34) $$
whence using (4.28 b ) to eliminate $ \tilde v_{N-1} $ we get 
$$ e^{-v_i } =\frac{ p_N (u_N - \tilde u_{N-1}) }{p_N ( q_N+\tilde u_{N-1} ) - c_N    } \;\; \frac{1 }{ ( u_i -c_N ) ^2 }  \;\;\frac{\prod_{ m = 1  }^{N-1} ( \tilde u_i - c_m )^2}{  \prod_{ j = 1  }^{N-2} ( u_i - \tilde u_j )   }, \;\;\;\;\; i=1,2,....,N-1. \eqno (4.35)    $$
Thus the mapping $ X_N $ is formally defined by the following relations 
$$p_N ^ 2 (q_N + u_N ) ( q_N + \tilde u_{N-1} ) + p_N [ (q_N + \tilde u_{N-1}  )(\Sigma_{ i=1}^{N-1}  u_i -   c_N  ) - (q_N + u_N ) ( \Sigma_{j=1}^{N-2} \tilde u_j +   c_N )  $$ $$+ 2\Sigma _{i=1}^{N-1} c_i    ( u_N - \tilde u_{N-1} )] + c_N [ c_N +\Sigma_{j=1}^{N-2} \tilde u_j -\Sigma_{i=1}^{N-1}   u_i ] =0 \eqno ( 4.36)  $$
$$ e^{-v_i } =\frac{ p_N (u_N - \tilde u_{N-1}) }{p_N ( q_N+ u_N ) - c_N    } \;\; \frac{1 }{ ( u_i -c_N ) ^2 }  \;\;\frac{\prod_{ m = 1  }^{N-1} ( \tilde u_i - c_m )^2}{  \prod_{ j = 1  }^{N-2} ( u_i - \tilde u_j )   }, \;\;\;\;\; i=1,2,....,N-1 \eqno (4.37) $$
and 
$$ e^{-\tilde v_j } = - \frac{ p_N (u_N - \tilde u_{N-1}) }{p_N ( q_N+\tilde u_{N-1} ) - c_N    } \prod_{ k = 1  }^{N-1} \frac{ ( \tilde u_j - c_k )^2}{ (\tilde u_j -u_k) } \;\;\;\; j=1,2,............., N-2 \eqno (4.38) $$
together with the relation 
 $$   v_N = \frac{ p_N (\tilde u_{N-1} + q_N) -c_N }{ u_N  - \tilde u_{n-1}},\;\;\;\;     \tilde v_{N-1} = \frac{ p_N (u_N + q_N ) -c_N  }{u_N  - \tilde u_{n-1} }  \eqno( 4.39 )  $$
  For the special case when $ c_i = 0 , \;\;\;for \; all i=1,2,......,N $ these expressions reduce to the following
$$ p_N = [ \frac{ \Sigma_{j=1}^{N-2} \tilde u_j}{ q_N +\tilde u_{N-1} } - \frac { \Sigma_{i=1}^{N-1}   u_i }{ q_N+u_N } ] , \;\;  v_N = \frac{ p_N (\tilde u_{N-1} + q_N)   }{ u_N  - \tilde u_{N-1}},\;\;  \tilde v_{N-1} = \frac{ p_N (u_N + q_N ) }{ u_N  - \tilde u_{N-1} } \eqno( 4.40a, b, c )   $$ 
$$e^{-v_i } =\frac{   u_N - \tilde u_{N-1}  }{  ( q_N+ u_N )      }  \; \frac{ u_i^2 } {\prod_{ j = 1  }^{N-2} ( u_i - \tilde u_j )   } \;\; \;\; i=1,2,.................... , N-1 .\eqno( 4.40d) $$
$$e^{-\tilde v_j } = - \frac{  u_N - \tilde u_{N-1} }{ q_N+\tilde u_{N-1}} \prod_{ k = 1  }^{N-1} \frac{ ( \tilde u_j ^2}{ (\tilde u_j -u_k) } \;\;\; j= 1,2,.............................., N-2. \eqno( 4.40e) $$
It is easy to verify that the expressions given in equations (4.40 a,b,c,d,e ) are derivable from the following generating function  $ F ( q_N ; \vec u ,\vec  { \tilde  u }) $, by means  of the following mapping $ X_N $ : 
$$ v_i = \frac { \partial F }{ \partial u_i } \;\;\;\; i=1,2,.............., N . \hskip 175pt$$
$$  \tilde v_j = - \frac { \partial F }{ \partial \tilde u_j} \;\;\;\; j=1,2,.............., N-1.\hskip 150pt  $$    
$$ p_N = - \frac { \partial F }{ \partial q_N  }  .\hskip 250pt \eqno( 4.41 ) $$
 with
\\$$  F ( q_N ; \vec u ,\vec  { \tilde  u }) = [-\Sigma _{ j=1}^{N-2} \tilde u_j ] log ( q_N + \tilde u_{N-1} ) + [\Sigma _{ i=1}^{N-1}   u_i ] log ( q_N + u_N ) $$$$+ [ \Sigma _{ j=1}^{N-2} \tilde u_j - \Sigma _{ i=1}^{N-1} u_i ]  log ( u_N-\tilde u_{N-1} )  $$$$+ 2 (N-1) \Sigma _{ j=1}^{N-2} \Theta (\tilde u_j) -2(N-2) \Sigma _{ i=1}^{N-1} \Theta (\tilde u_j) $$$$ +    \Sigma _{ i=1}^{N-1} \Sigma _{ j=1}^{N-2} \Theta (u_i - \tilde u_j) + i \pi ( N-2 ) \Sigma _{ j=1}^{N-2} \tilde u_j     \eqno ( 4.42 ) $$
 where $ \Theta  ( u ) = \int log (u ) du   + c ( constant ). $
 The generating function for the subsequent mapping $ X_{N-1} $ is obtained by replacing  $ N $ by $ N-1 $ and so on. The above formulae thus describe explicitly the inverse separation map $ S_N^{-1} $ and by virtue  of the existence of a generating function for each stage we conclude that the inverse separation map is canonical.
\section {  Discussion}   In this communication we have studied the nature of the inverse  separation transformation between the local nonlinear variables of  a discrete integrable system, and the corresponding separation variables of the system. The novel feature of the methodology used is the fact that this separation transformation can be factorized into a sequence of transformations, each with its corresponding generating function. The latter clearly is indicative of the canonical nature of the separation transformation. 
\section{ Acknowledgement } One of the authors (S.M),  is grateful to UGC for a Project Associate fellowship which made this work possible. \newpage  \section { References }
 1. V.B. Kuznetsov, M. Salerno, E.K.Sklyanin  {\it J.Phys. A.Math Gen.}{\bf  33A} (2000),  171  \\
2. V. B. Kuznetsov, M.Petrero, O.Ragnisco, { \it nlin.SI/0403028} (2004) \\
3. L.Faddeev and L. Takhtajan, {\it Hamiltonian methods in the theory of solitons}, Springer Verlag, Berlin, (1988)\\
4. I.V.Komarov, { \it Theo. Mat. Fiz.}, {\bf 50 } (3), 402, (1982).   {\it J.Phys.}{\bf A15 }(6) (1765) 1982\\
5. E. K. Sklyanin  {\it  Prog. Theo. Phys. Suppl.} {\bf 118 }, 35, (1995)\\
6. V. B. Kuznetsov -{\it Inverse problem for $sl(2)$ lattices} nlin.SI/ 0207025\\
7. K.Takasaki, {\it Spectral curve, Darboux coordinates and Hamiltonian structure of periodic  dressing chain}, {\it nlin.SI/0206049}\\
8. E. K. Sklyanin, {\it Differential Geometry, Lie groups and Mechanics, } VI Zap. Nauchn. Sem. Leningrad, (LOMI) {\bf 133}, 236, (1984).\\
9. V. B. Kuznetsov, {\it Kowalevski top revisited}, SI/0110012.

\end{document}